\documentclass[a4paper,11pt]{article}
\pdfoutput=1 

\usepackage{jheppub} 

\usepackage[T1]{fontenc} 

\usepackage{tensor}
\usepackage{graphicx,subcaption}
\graphicspath{ {figures/} }	
\usepackage{amsfonts}
\usepackage{xparse}
 \usepackage{amsmath}
 \usepackage{amssymb}
 \usepackage{mathtools}
\usepackage{epsfig}
\usepackage{pdfpages}
\usepackage{bbm}
\usepackage{graphicx,epstopdf}
\usepackage[numbers]{natbib}
\usepackage[makeroom]{cancel}
\usepackage{hyperref}
\usepackage{array}
\usepackage[export]{adjustbox}
\usepackage[normalem]{ulem}
\usepackage{tcolorbox}
\usepackage{romannum}

\newcommand{\eq}[1]{(\ref{#1})}
\newcommand{\fig}[1]{Fig. \ref{#1}}

\newcommand{\nn}{\nonumber}

\def\T{\mathcal{T}} \def\O{\mathcal{O}} \def\S{\mathcal{S}}\def\J{\mathcal{J}}\def\V{\mathcal{V}}\def\Q{\mathcal{Q}}\def\F{\mathcal{F}}

\title{Holographic thermodynamics of rotating black holes}

\author[a]{Ting-Feng Gong,}
\author[b]{Jie Jiang,}
\author[a,1]{Ming Zhang \note{Corresponding author}}

\affiliation[a]{Department of Physics, Jiangxi Normal University,\\ Nanchang 330022, China}
\affiliation[b]{College of Education for the Future, Beijing Normal University, \\ Zhuhai 519087, China}

\emailAdd{tingfenggong@jxnu.edu.cn}
\emailAdd{jiejiang@mail.bnu.edu.cn}
\emailAdd{mingzhang@jxnu.edu.cn}

\abstract{
We provide mass/energy formulas for the extended thermodynamics, mixed thermodynamics, and holographic conformal field theory (CFT) thermodynamics for the charged and rotating Kerr-Newman Anti-de Sitter black holes. Then for the CFT thermal states dual to the black hole, we find the first-order phase transitions and criticality phenomena in the canonical ensemble with fixed angular momentum, volume, and central charge. We observe that the CFT states cannot be analogous to the Van der Waals fluids, despite the critical exponents falling into the universality class predicted by the mean field theory. Additionally,  we examine the (de)confinement phase transitions within the grand canonical ensemble with fixed angular velocity, volume, and central charge of the CFT. Our findings suggest that the near zero temperature (de)confinement phase transitions can occur with the angular velocity of the CFT that solely depends on the CFT volume.
}

\begin{document}

\maketitle
\flushbottom

\section{Introduction}

Thermodynamics of black holes has advanced significantly since the proposal of the four laws of black hole thermodynamics in the 1970s \cite{Bardeen:1973gs, Bekenstein:1973ur}. In particular, significant attention has been devoted to the thermodynamics of asymptotically Anti-de Sitter (AdS) black holes, which can be interpreted dually as thermal states within the conformal field theory (CFT) via the AdS/CFT correspondence \cite{Maldacena:1997re}. The Hawking-Page transition, a first-order transition between a large black hole and thermal radiation in the bulk AdS spacetime, is one example that has garnered significant interest in this regard, along with its corresponding thermalization transition in the boundary strongly coupled dual CFT \cite{Hawking:1982dh, Witten:1998zw}.

In the realm of {\it extended thermodynamic phase space}, the AdS black hole's phase structure and transitions have been significantly enriched by treating the negative cosmological constant $\Lambda$ as a thermodynamic pressure \cite{Kastor:2009wy,Dolan:2010ha,Dolan:2011xt,Cvetic:2010jb}. One well-known example is the Van der Waals-like phase transition of charged AdS black holes \cite{Chamblin:1999tk,Chamblin:1999hg,Cvetic:1999ne,Kubiznak:2012wp}, along with that of rotating AdS black holes \cite{Gunasekaran:2012dq,Wei:2015ana,Cheng:2016bpx}. Black hole chemistry \cite{Kubiznak:2016qmn} has opened doors for a broad range of thermodynamic phenomena, such as superfluid behavior \cite{Hennigar:2016xwd}, reentrant phase transitions \cite{Frassino:2014pha}, possible interpretation of black hole microstructures \cite{Wei:2019uqg}, and multicritical phase transitions \cite{Tavakoli:2022kmo,Wu:2022plw}. The variable cosmological constant has long puzzled researchers, however, recent work \cite{Frassino:2022zaz} has provided an answer by utilizing holographic braneworlds as higher-dimensional origins for extended black hole thermodynamics.

The holographic interpretation of black hole chemistry is a topic that has attracted significant attention in recent years \cite{Kastor:2014dra,Dolan:2014cja,Zhang:2014uoa}. This interpretation proposes that the thermodynamics of black holes in the bulk is equivalent to the thermodynamics of strongly coupled gauge theories on the boundary, given a large number of degrees of freedom in the limit of a large $N$. In the past, variations in the bulk cosmological constant have caused changes in both the central charge of the conformal field theory (CFT) (or the number of colors $N$) and the boundary volume $\mathcal{V}$ \cite{Karch:2015rpa}. Additionally, the CFT electric charge and its conjugate chemical potential are related to the AdS length scale. Therefore, the thermodynamic first law of the bulk spacetime cannot be directly projected onto the holographic dual boundary field theory. To address this issue, researchers have recently proposed a theory in which Newton's constant is dynamic to maintain the CFT's central charge fixed \cite{Visser:2021eqk, Cong:2021fnf, Cong:2021jgb}.

In this new paradigm, the dual CFT remains unchanged even when the bulk cosmological constant is rescaled. As a consequence, we can derive the bulk first law of black hole thermodynamics from the first law of the dual boundary field theory.  On one hand, this approach provides a holographic interpretation for the black hole chemistry, specifically for the Van der Waals phase transition of the charged Reissner-Nordström (RN) AdS black hole \cite{Cong:2021fnf}. This transition is determined by the degrees of freedom of its dual field theory in the large $N$ limit, as confirmed by central charge criticality investigations for non-linear electromagnetic black holes \cite{Kumar:2022fyq,Bai:2022vmx} and Gauss-Bonnet black holes \cite{Kumar:2022afq,Qu:2022nrt}. On the other hand, we can study the phase behaviors in the holographic thermodynamics of charged AdS black holes \cite{Cong:2021jgb}. The CFT description of these black holes exhibits zeroth-order, first-order, and second-order phase transitions. However, it was found that there is no pressure-volume criticality in the CFT, which verifies that the CFT state dual to the charged AdS black hole is not a Van der Waals fluid.

The purpose of this research paper is to generalize previous investigations of holographic thermodynamics. There are two main aspects that we will be focusing on. Firstly, we will propose generalized mass/energy formulas for charged and rotating Kerr-Newman-AdS (KN-AdS) black holes, on both the bulk AdS side and on the holographic CFT side. Secondly, we will generalize the holographic thermodynamics study for charged RN-AdS black holes, as previously studied in \cite{Cong:2021jgb}, to rotating Kerr-AdS black holes. We aim to explore the criticality and (de)confinement phase transitions of the CFT states, which are dual to rotating AdS black holes and exist in a rotating Einstein universe.  In section \ref{mass}, we will focus on generalized mass/energy formulas of charged rotating KN-AdS black holes in extended thermodynamics, mixed thermodynamics, and CFT thermodynamics. In section \ref{caen}, we will delve into the phase transitions and criticality of the CFT thermodynamics for rotating AdS black holes in the canonical ensemble, with fixed angular momentum, volume, and central charge. Afterward, we will explore the (de)confinement phase behaviors of the black holes in the grand canonical ensemble with fixed angular velocity, volume, and central charge in section \ref{oten}.  We will present concluding remarks in the last section. Please refer to our previous study on the topological property of bulk/boundary thermodynamics for charged RN-AdS black holes in \cite{Zhang:2023uay}.

\section{Generalized mass/energy formulas}\label{mass}
In what follows, we will first give generalized mass formulas for the charged and rotating KN-AdS black holes in the extended thermodynamics, in terms of bulk event horizon area, electric charge, angular momentum, and cosmological constant. Then we will further propose the generalized mass formulas for the black holes in the mixed thermodynamics, in terms of  black hole entropy, electric charge, angular momentum, thermodynamic pressure, and boundary CFT central charge. After that, we will propose an energy formula for the holographic dual field theory of the KN-AdS black hole.

\subsection{Extended thermodynamics}
The action of the Einstein-Maxwell theory we consider in this paper is \footnote{Note that a standard form of the action reads \begin{equation}
I=\frac{1}{16 \pi G_N} \int d^{4} x \sqrt{-g}\left(R-2 \Lambda-G_N \F^2\right).
\end{equation} However, the convention \eqref{actb} we use here (see also in \cite{Chamblin:1999tk}) will not change the physics.}
\begin{equation}\label{actb}
I=\frac{1}{16 \pi G_N} \int d^{4} x \sqrt{-g}\left(R-2 \Lambda-\F^2\right),
\end{equation}
where $G_N$ is the Newton's constant, $\Lambda$ is the cosmological constant, $\F_{ab}$ is the strength of the $U(1)$ field, $R$ is the Ricci scalar. The corresponding  solution of the field theory is
\begin{equation}\label{le}
ds^{2}=-\frac{\Delta_{r}}{\Sigma^2} \left(dt- \frac{a \sin^2\theta}{\Xi} d\phi \right)^2 +\Sigma^2  \left(\frac{dr^2}{\Delta_{r}}+\frac{d\theta^2}{H_\theta} \right)+ \frac{H_\theta \sin^2\theta}{\Sigma^2} \left[ a dt -\frac{r^2+a^2}{\Xi}d\phi\right]^2,
\end{equation}
\begin{gather}
\F_{ab}=(dB)_{ab},\\ \,B=-\frac{G_N qr}{\Sigma^2}\left[dt-a\sin^2\theta d\phi\right]+\Phi_t dt,
\end{gather}
where
\begin{eqnarray}
   \Delta_{r}&=&r^2-2G_N m r+a^2+G_N^2 q^2+\frac{(r^2+a^2)r^2}{l^2},\nn\\
   H_{\theta}&=&1-\frac{a^2}{l^2}\cos^2\theta,\nn\\
   \Sigma^2&=&r^{2}+a^{2}\cos^{2}\theta,\nn\\
   \Phi_t &=&\frac{G_N qr_+}{a^2+r_+^2},\nn\\
   \Xi&=&1-\frac{a^2}{l^2}.
\end{eqnarray}
Here the event horizon $r_+$ of the black hole is given by $\Delta_r=0$. $B_a$ is the gauge potential where the given value of $\Phi_t$ makes the gauge potential on the event horizon of the black hole vanish. $m, q, a$ are parameters related to the ADM mass, electric charge, and angular momentum of the black hole via the relation
\begin{equation}\label{main}
M=\frac{m}{\Xi^2},
\end{equation}
\begin{equation}\label{mjq}
 J=\frac{a m}{\Xi^2}=Ma,\quad Q=\frac{q}{\Xi}.
\end{equation}
$l$ is the AdS length scale or the curvature radius of the AdS spacetime, related to the cosmological constant by
\begin{equation}\label{lambl}
\Lambda=-\frac{3}{l^2}.
\end{equation}

The horizon area, temperature,  angular velocity, thermodynamic pressure, thermodynamic volume, and electric potential of the black hole respectively are
\begin{equation}
A=\frac{4 \pi\left(r_{+}^2+a^2\right)}{\Xi},
\end{equation}
\begin{equation}
T=\frac{-l^2 \left(a^2+G_N^2 q^2\right)+r_+^2 \left(a^2+l^2\right)+3 r_+^4}{4 \pi  l^2 r_+ \left(a^2+r_+^2\right)},
\end{equation}
\begin{equation}
\Omega_b=\frac{a \left(l^2+r_+^2\right)}{l^2 \left(a^2+r_+^2\right)},
\end{equation}
\begin{equation}\label{preb}
P=-\frac{\Lambda}{8\pi G_N},
\end{equation}
\begin{equation}
V=\frac{2 \pi  l^2 \left(a^2 l^2 \left(a^2+G_N^2 q^2\right)-\left(r_+^4 \left(a^2-2 l^2\right)\right)-r_+^2 \left(a^4-3 a^2 l^2\right)\right)}{3 r_+ \left(a^2-l^2\right)^2},
\end{equation}
\begin{equation}
\Phi=\frac{G_N qr_+}{a^2+r_+^2}.
\end{equation}
The entropy $S$ and surface gravity $\kappa$ of the black hole are given by
\begin{align}\label{entb}
S&=\frac{A}{4G_N},\\
\kappa&=2\pi T.\label{kapt}
\end{align}

In \cite{Caldarelli:1999xj}, a generalized mass formula for the KN-AdS black hole was proposed, where the Newton's constant was set to 1. Here we restore the Newton's constant in the mass formula and obtain
\begin{equation}\label{mass1}
M^2=\frac{16 G_N^3 P^2 S^3}{9 \pi }+\frac{4}{3} G_N^2 P Q^2 S+\frac{8}{3} \pi  G_N J^2 P+\frac{\pi  J^2}{G_N S}+\frac{4 G_N P S^2}{3 \pi }+\frac{\pi  G_N Q^4}{4 S}+\frac{S}{4 \pi  G_N}+\frac{Q^2}{2}.
\end{equation}
By this mass formula, we can write the thermodynamic first law for the KN-AdS black hole in the thermodynamic extended phase space as
\begin{equation}\label{beps}
dM=TdS+\Phi dQ+VdP+\Omega_b dJ.
\end{equation}
The corresponding integral Smarr formula is
\begin{equation}\label{bsm}
M=2TS+2\Omega_b J+\Phi Q-2PV.
\end{equation}

In order to incorporate the variation of the Newton's constant $G_N$, we can release the dependence of the entropy and pressure on the Newton's constant with the aid of \eq{preb} and \eq{entb}. Then we have another form of the mass formula for the black hole as
\begin{equation}\label{mass1t}
M^2=\frac{A^3 \Lambda ^2}{2304 \pi ^3 G_N^2}-\frac{A^2 \Lambda }{96 \pi ^2 G_N^2}+\frac{\pi  G_N^2 Q^4}{A}+\frac{A}{16 \pi  G_N^2}+\frac{4 \pi  J^2}{A}-\frac{A \Lambda  Q^2}{24 \pi }-\frac{J^2 \Lambda }{3}+\frac{Q^2}{2}.
\end{equation}
Thus we can have the thermodynamic first law of the black hole which treats $\Lambda, G_N$ as dynamic parameters, 
\begin{equation}\label{fir2}
d M=\frac{\kappa}{8 \pi G_N} d A+\Phi d Q+\Omega_b dJ-\frac{V}{8 \pi G_N} d \Lambda-(M-\Phi Q-\Omega_b J) \frac{d G_N}{G_N},
\end{equation}
Related quantities in this law can be derived directly as
\begin{equation}
\begin{aligned}
\frac{\kappa}{8 \pi G_N}&=\left(\frac{\partial M}{\partial A}\right)_{Q, J, \Lambda, G_N}\\&=\frac{1}{2M}\left(\frac{A^2 \Lambda ^2}{768 \pi ^3 G_N^2}-\frac{\pi  G_N^2 Q^4}{A^2}-\frac{4 \pi  J^2}{A^2}-\frac{A \Lambda }{48 \pi ^2 G_N^2}+\frac{1}{16 \pi  G_N^2}-\frac{\Lambda  Q^2}{24 \pi }\right),
\end{aligned}
\end{equation}
\begin{equation}
\Phi=\left(\frac{\partial M}{\partial Q}\right)_{A, J, \Lambda, G_N}=\frac{1}{2M}\left(\frac{4 \pi  G_N^2 Q^3}{A}-\frac{A \Lambda  Q}{12 \pi }+Q\right),
\end{equation}
\begin{equation}
-\frac{V}{8 \pi G_N}=\left(\frac{\partial M}{\partial \Lambda}\right)_{A, Q, J, G_N}=\frac{1}{2M}\left(\frac{A^3 \Lambda }{1152 \pi ^3 G_N^2}-\frac{A^2}{96 \pi ^2 G_N^2}-\frac{A Q^2}{24 \pi }-\frac{J^2}{3}\right),
\end{equation}
\begin{equation}
\Omega_b=\left(\frac{\partial M}{\partial J}\right)_{A, Q, \Lambda}=\frac{1}{2M}\left(\frac{8 \pi  J}{A}-\frac{2 J \Lambda }{3}\right),
\end{equation}
\begin{equation}
\begin{aligned}
\frac{\Omega_b J+\Phi Q-M}{G_N}&=\left(\frac{\partial M}{\partial G_N}\right)_{A, Q, J, \Lambda}\\&=\frac{1}{2M}\left(-\frac{A^3 \Lambda ^2}{1152 \pi ^3 G_N^3}+\frac{A^2 \Lambda }{48 \pi ^2 G_N^3}-\frac{A}{8 \pi  G_N^3}+\frac{2 \pi  G_N Q^4}{A}\right).
\end{aligned}
\end{equation}
In the vanishing angular momentum $J$ limit, the results reduce to the ones for the charged Reissner-Nordström AdS black hole in \cite{Cong:2021jgb} and also in \cite{Zhang:2023uay}. One can also check these relations by using \eq{main}-\eq{kapt}. In this treatment, the  integral Smarr formula corresponding to the first law \eq{fir2} reads
\begin{equation}
M=\frac{\kappa A}{4 \pi G_N}+2\Omega_b J+\Phi Q+\frac{V \Lambda}{4 \pi G_N},
\end{equation}
which is equivalent to \eq{bsm}.

\subsection{Mixed thermodynamics}
Now we want to transfer the bulk thermodynamic first law \eq{fir2} into a form containing the boundary central charge. We employ the holographic dual relation between the central charge $C$, the bulk AdS scale $l$, and the Newton's constant $G_N$ in the Einstein gravity,
\begin{equation}\label{cogn}
C=\frac{\Omega_{2} l^{2}}{16 \pi G_N},
\end{equation}
where $\Omega_2$ is the volume of a 2-sphere. Combing this relation with  \eq{lambl} and \eq{preb}, we can get the Newton's constant $G_N$ in terms of the boundary central charge and the bulk thermodynamic pressure as
\begin{equation}
G_N=\frac{1}{4} \sqrt{\frac{3}{2 \pi  C P}}.
\end{equation}
Then we can substitute this relation into \eq{mass1}, resulting in a mass formula for the KN-AdS black hole encoding the  boundary central charge as
\begin{equation}\label{mass2}
\begin{aligned}
M^2=&\frac{4 \pi ^{3/2} \sqrt{\frac{2}{3}} \sqrt{CP} J^2 }{S}+\frac{\sqrt{\frac{2 \pi }{3}} J^2 \sqrt{P}}{\sqrt{C}}+\frac{\sqrt{\frac{3 \pi }{2}} Q^4}{16 \sqrt{CP}  S}+\frac{Q^2}{2}\\&+\frac{\sqrt{P} S^3}{8 \sqrt{6} \pi ^{5/2} C^{3/2}}+\frac{\sqrt{P} S^2}{\sqrt{6} \pi ^{3/2} \sqrt{C}}+\sqrt{\frac{2}{3 \pi }} \sqrt{CP}  S+\frac{Q^2 S}{8 \pi  C}.
\end{aligned}
\end{equation}

This mass formula admits a thermodynamic first law in a bulk-boundary mixed form as
\begin{equation}\label{mifir}
d M=T d S+\Phi d Q+V_C d P+\mu_b d C+\Omega_b dJ,
\end{equation}
where the redefined thermodynamic volume  $V_C$ and the chemical potential  $\mu_b$ conjugate respectively to the pressure and the central charge emerge.  Thermodynamic quantities in this mixed first law can be obtained as
\begin{equation}
\begin{aligned}
T=\left(\frac{\partial M}{\partial S}\right)_{Q, P, C, J}=&\frac{1}{2M}\left(\frac{\sqrt{\frac{3}{2}} \sqrt{P} S^2}{8 \pi ^{5/2} C^{3/2}}-\frac{4 \sqrt{\frac{2}{3}} \pi ^{3/2} \sqrt{CP} J^2 }{S^2}-\frac{\sqrt{\frac{3 \pi }{2}} Q^4}{16 \sqrt{CP}  S^2}\right.\\&\left.+\frac{\sqrt{\frac{2}{3}} \sqrt{P} S}{\pi ^{3/2} \sqrt{C}}+\sqrt{\frac{2}{3 \pi }} \sqrt{CP} +\frac{Q^2}{8 \pi  C}\right),
\end{aligned}
\end{equation}
\begin{equation}
\Phi=\left(\frac{\partial M}{\partial Q}\right)_{S, P, C, J}=\frac{Q \left(8 \pi  C \sqrt{P} S+\pi ^{3/2} \sqrt{6} \sqrt{C} Q^2+2 \sqrt{P} S^2\right)}{16 \pi  C M \sqrt{P} S},
\end{equation}
\begin{equation}
\begin{aligned}
V_C&=\left(\frac{\partial M}{\partial P}\right)_{S, Q, C, J}\\&=\frac{32 \pi ^2 C^2 P \left(4 \pi ^2 J^2+S^2\right)+\pi ^3 C \left(32 J^2 P S-3 Q^4\right)+16 \pi  C P S^3+2 P S^4}{64 \sqrt{6} \pi ^{5/2} C^{3/2} M P^{3/2} S},
\end{aligned}
\end{equation}
\begin{equation}
\begin{aligned}
\mu=&\left(\frac{\partial M}{\partial C}\right)_{S, Q, P, J}\\=&\frac{1}{384 \pi ^{5/2} C^{5/2} M \sqrt{P} S}\left(32 \pi ^2 \sqrt{6} C^2 P \left(4 \pi ^2 J^2+S^2\right)\right.\\&\left.-\sqrt{6} \pi  C \left(\pi ^2 \left(32 J^2 P S+3 Q^4\right)+16 P S^3\right)-24 \pi ^{3/2} \sqrt{CP}  Q^2 S^2-6 \sqrt{6} P S^4\right),
\end{aligned}
\end{equation}
\begin{equation}
\Omega_b=\left(\frac{\partial M}{\partial J}\right)_{S, Q, P, C}=\frac{\sqrt{\frac{2 \pi }{3}} J \sqrt{P} (4 \pi  C+S)}{\sqrt{C} M S}.
\end{equation}

\begin{figure}[tbp!]
\begin{center}
\includegraphics[width=90mm,angle=0]{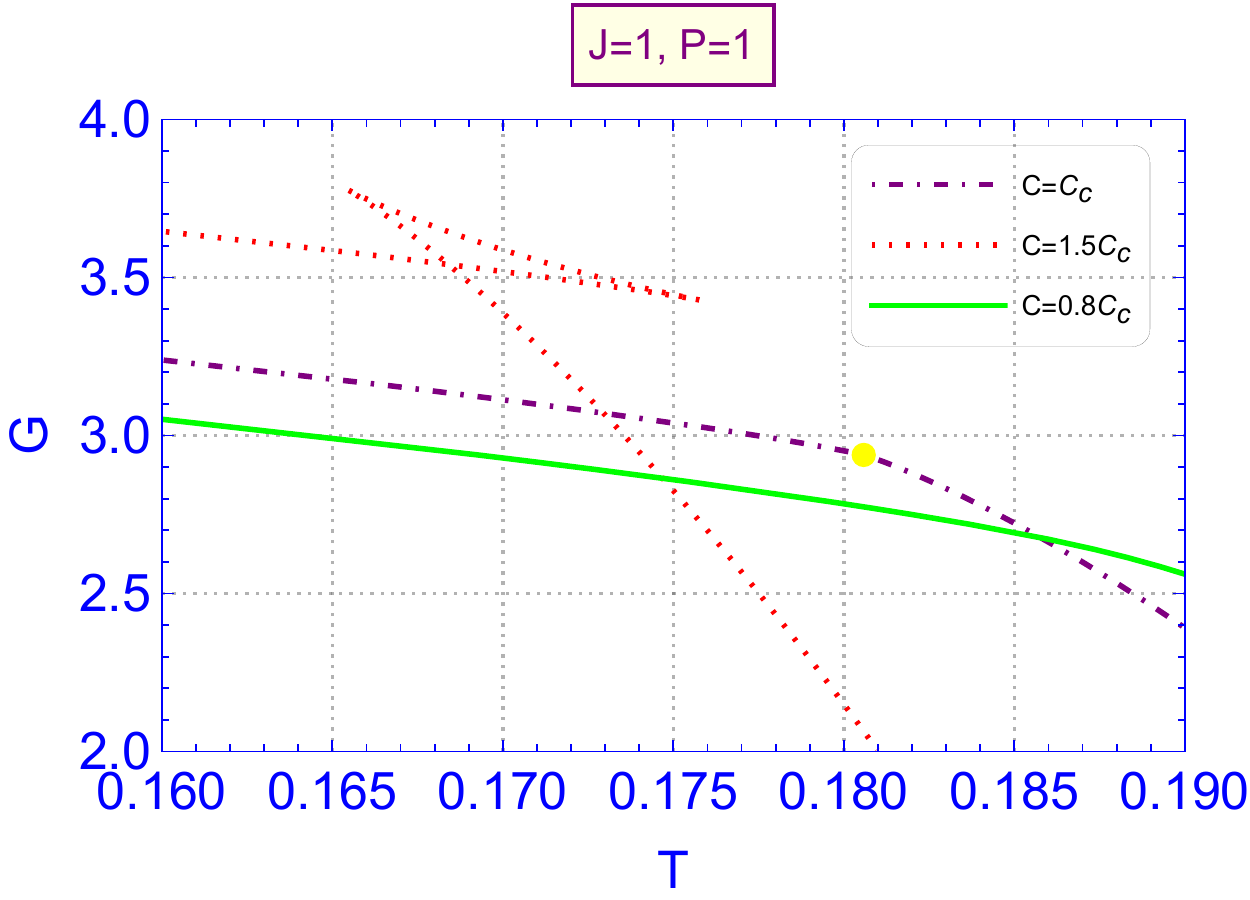}
\end{center}
\vspace{-5mm}
 \caption {Variations of the free energy with respect to the temperature for the Kerr-AdS black hole with the critical central charge $C_c=10.4459$. The black hole takes a second-order phase transition at the  yellow point. }
 \label{fig0}
\end{figure}

Thermodynamics of the Kerr-AdS black hole in the setting of a dynamic cosmological constant has been studied extensively in  \cite{Gunasekaran:2012dq,Altamirano:2013uqa,Altamirano:2013ane,Wei:2015ana,Yang:2021ljn}. Here we show the central charge criticality of the rotating black hole in the thermodynamic extended phase space with dynamic Newton's constant and cosmological constant. 

We can find the Van der Waals-like phase transition and criticality of the rotating Kerr-AdS black hole in the fixed $(J, P, C)$ ensemble. The Gibbs free energy in this ensemble reads
\begin{equation}
G=M-TS=\frac{\sqrt{PS} \left(16 \pi ^2 C^2 \left(12 \pi ^2 J^2+S^2\right)+32 \pi ^3 C J^2 S-S^4\right)}{4\ 2^{3/4} \sqrt[4]{3} \pi ^{5/4} C^{3/4} S \sqrt{\sqrt{P} (4 \pi  C+S) \left(4 \pi  C \left(4 \pi ^2 J^2+S^2\right)+S^3\right)}}.
\end{equation}
\fig{fig0} depicts the central charge criticality of  the rotating Kerr-AdS black hole. When the central charge exceeds the critical value ($C>C_c$), the black hole undergoes a first-order phase transition between small and large black holes. At the critical point ($C=C_c$), there is a second-order phase transition with no latent heat. In the regime where the central charge is less than the critical value ($C<C_c$), there is no distinct phase of the black hole. 

To get the analytic critical point of the phase transition, we content ourselves to set $Q=0$ and expand the temperature $T$ in the small regime of $J$ to the order of $\O(J^2)$. Thus we have
\begin{equation}
T=\frac{\sqrt[4]{P} \left(4 \pi  C \left(S^2-6 \pi ^2 J^2\right)+3 S^3\right)}{4\ 2^{3/4} \sqrt[4]{3} \pi ^{5/4} C^{3/4} S^{5/2}}+\O(J^2).
\end{equation}
Then by the restriction equations
\begin{equation}\label{crit}
\left(\frac{\partial T}{\partial S}\right)_{C, J}=0=\left(\frac{\partial^2 T}{\partial S^2}\right)_{C, J},
\end{equation}
we get the critical point for the Van der Waals-like phase transition of the Kerr-AdS black hole in the central charge criticality as
\begin{equation}
S_c= 3 \sqrt{10} \pi  J,\quad C_c=\frac{27}{4} \sqrt{\frac{5}{2}} J.
\end{equation}

\subsection{CFT thermodynamics}

We are now to construct the mass formula for the boundary CFT thermodynamics of the charged and rotating KN-AdS black holes. We set the metric of the CFT as \cite{Gubser:1998bc,Witten:1998qj,Ahmed:2023snm}
\begin{equation}
d s^2=\omega^2 \left(-d t^2+l^2 d \Omega_{2}^2\right),
\end{equation}
where $\omega$ is an arbitrary dimensionless conformal factor, $d \Omega_{2}^2$ is the line element of a 2-sphere. In 
\cite{Visser:2021eqk,Cong:2021jgb}, the conformal factor is chosen to be $\omega=R/l$, with $R$ the curvature radius of the boundary, but here we allow it to be a generic parameter, as has been advised recently in \cite{Ahmed:2023snm} ( if we take $\omega=R/l$, we will certainly obtain the same results shown below). Definitely, as pointed out in \cite{Ahmed:2023snm}, the variable $\omega$ then makes us possible to formulate a holographic first law which is exactly dual to the first law of extended black hole thermodynamics \eq{beps} with fixed Newton's constant $G_N$ and variable cosmological constant $\Lambda$.

The spatial volume of the boundary sphere is proportional to $(\omega l)^2$, for definiteness, we choose it to be \footnote{Other selection of the proportional constant will only rescale the overall energy by a constant factor, i.e., if we take $\Omega_2\to k\Omega_2$ with $k$ a constant, then we will have $E^2\to kE^2$ in \eq{mass3}.}
\begin{equation}\label{voml}
\mathcal{V}=\Omega_{2} (\omega l)^{2}.
\end{equation}
Associated with the CFT volume $\V$, a conjugate CFT pressure $p$ should be presented, so that there is a work term $-pd\V$. The holographic dictionary relating bulk mass, temperature, entropy, angular momentum, electric potential, and charge with their boundary counterparts reads \cite{Chamblin:1999tk,Savonije:2001nd,Visser:2021eqk}
\begin{equation}
E=\frac{M}{\omega},\quad  \mathcal{T}=\frac{T}{\omega},\quad \mathcal{S}=S,\quad \mathcal{J}=J,\quad \varphi=\frac{\Phi}{\omega l}, \quad \mathcal{Q}=Q l.
\end{equation}

Then employing  \eq{cogn}, \eq{mass2}, and \eq{voml}, we can obtain an energy formula for the boundary CFT written as
\begin{equation}\label{mass3}
E^2=\frac{16 \pi ^2 C^2 \left(4 \pi ^2 \mathcal{J}^2+\mathcal{S}^2\right)+8 \pi  C \mathcal{S} \left(\pi ^2 \left(2 \mathcal{J}^2+\mathcal{Q}^2\right)+\mathcal{S}^2\right)+\left(\pi ^2 \mathcal{Q}^2+\mathcal{S}^2\right)^2}{4 \pi ^2 C \mathcal{S} \mathcal{V}}.
\end{equation}
This internal energy formula admits the first law and the Euler relation for the boundary CFT  as
\begin{align}
d E&=\mathcal{T} d \mathcal{S}+\varphi d \mathcal{Q}-p d \mathcal{V}+\mu d C+\Omega d\mathcal{J}\label{affir},\\
E&=\mathcal{T} \mathcal{S}+\varphi \mathcal{Q}+\mu C+\Omega \mathcal{J},
\end{align}
where related quantities are specifically given by
\begin{equation}\label{cfttemp}
\mathcal{T}=\left(\frac{\partial E}{\partial \mathcal{S}}\right)_{\mathcal{Q}, \mathcal{V}, C}=\frac{16 \pi ^2 C^2 \left(\mathcal{S}^2-4 \pi ^2 \mathcal{J}^2\right)+16 \pi  C \mathcal{S}^3-\pi ^4 \mathcal{Q}^4+2 \pi ^2 \mathcal{Q}^2 \mathcal{S}^2+3 \mathcal{S}^4}{8 \pi ^2 C E \mathcal{S}^2 \mathcal{V}},
\end{equation}
\begin{equation}
\varphi=\left(\frac{\partial E}{\partial \mathcal{Q}}\right)_{\mathcal{S}, \mathcal{V}, C}=\frac{\mathcal{Q} \left(4 \pi  C \mathcal{S}+\pi ^2 \mathcal{Q}^2+\mathcal{S}^2\right)}{2 C E \mathcal{S} \mathcal{V}},
\end{equation}
\begin{equation}\label{eoseq}
-p=\left(\frac{\partial E}{\partial \mathcal{V}}\right)_{\mathcal{S}, \mathcal{Q}, C}=-\frac{E}{2\mathcal{V}},
\end{equation}
\begin{equation}
\mu=\left(\frac{\partial E}{\partial C}\right)_{\mathcal{S}, \mathcal{Q}, \mathcal{V}}=-\frac{\left(\pi ^2 \mathcal{Q}^2+\mathcal{S}^2\right)^2-16 \pi ^2 C^2 \left(4 \pi ^2 \mathcal{J}^2+\mathcal{S}^2\right)}{8 \pi ^2 C^2 E \mathcal{S} \mathcal{V}},
\end{equation}
\begin{equation}
\Omega=\left(\frac{\partial E}{\partial \J}\right)_{\mathcal{S}, \mathcal{Q}, \mathcal{V}}=\frac{4 \pi  \mathcal{J} (4 \pi  C+\mathcal{S})}{E \mathcal{S} \mathcal{V}}.
\end{equation}
We can see that the Euler relation respects the scaling property of the gauge field theory in the large-$N$ 't Hooft limit  at finite temperature \cite{tHooft:1973alw}.  \eq{eoseq} is the equation of state for the CFT. $\mu$ is the chemical potential conjugate to the central charge in the boundary.

We here want to note that the internal energy formula \eq{mass3} for the CFT can also be obtained from the bulk mass formula \eq{mass1}. More interestingly, we find that the affiliated CFT first law \eq{affir} is exactly dual to the bulk first law of extended black hole thermodynamics \eq{beps}, where no dynamic Newton's constant is needed, only with a variable cosmological constant $l$. This is strictly compatible with the formulation given recently in \cite{Ahmed:2023snm}.

One may question whether Newton's constant is variable, but we argue that a dynamic role for it only makes sense within the context of the mixed first law \eq{mifir}, In this case,  there is a modification to the former volume term $V$ in \eq{beps} to be $V_C$. At the same time, with this consideration, we can view the Van der Waals-like phase transition of the black holes in fixed boundary theory (i.e., with fixed central charge $C$). However, the derivation of the internal energy formula as well as the boundary first law and Euler relation do not require a dynamic Newton's constant. Here one does not need to consider a reasonable description of the bulk phase transition and the point here is to construct exact dual relation between bulk and boundary first laws. A dynamic $\omega$, instead, takes on the duty of preventing the degeneracy of the volume $\V$ and central charge $C$. In this sense, we specify the holographic dual relation raised by  \cite{Ahmed:2023snm} for the self-consistent dual relation between bulk and boundary first law to the uniform bulk and boundary mass/energy formulas for the  KN-AdS black hole.

\section{Phase transition of CFT in canonical ensemble }\label{caen}

In \cite{Cong:2021jgb}, thermodynamic phase structures of the charged CFT states corresponding to  charged AdS black holes were studied. Here, to study the thermodynamic behavior of CFT states in the rotating Einstein universe, we will now consider the phase transition of the dual CFT states for the uncharged rotating Kerr-AdS black holes in the canonical ensemble with fixed angular momentum $\J$, volume $\V$, and central charge $C$, by turning off the charge of both the black hole and the CFT.  In this ensemble, the energy formula \eq{mass3} we gave above can be used to calculate related thermodynamic quantities directly.  We will see that the black hole exhibits a first-order phase transition from low entropy states to high entropy states and a second-order criticality phenomenon at certain parameter regimes. We will analyze the roles of the angular momentum $\J$ and central charge $C$ in the phase transition and calculate the critical exponents for the phase transition at the critical point.

\subsection{Phase transitions and critical points}
In the ensemble with fixed $(\J, \mathcal{V}, C)$, we have the Helmholtz free energy as a thermodynamic potential, as 
\begin{equation}
dF=d(E-\T\S)=-\S d \T+\Omega d \J-p d \mathcal{V}+\mu d C.
\end{equation}
We can know that in an isothermal thermodynamic process of this ensemble, the system tends to be at a state with lower Helmholtz free energy. Specifically, we have the free energy of the  CFT as
\begin{equation}\label{helfrc}
\begin{aligned}
F \equiv E-\T \S=\frac{16 \pi ^2 C^2 \left(12 \pi ^2 \mathcal{J}^2+\mathcal{S}^2\right)+32 \pi ^3 C \mathcal{J}^2 \mathcal{S}-\mathcal{S}^4}{4 \pi  \sqrt{C \mathcal{S} \mathcal{V} (4 \pi  C+\mathcal{S}) \left(4 \pi  C \left(4 \pi ^2 \mathcal{J}^2+\mathcal{S}^2\right)+\mathcal{S}^3\right)}}.
\end{aligned}
\end{equation}
In the small angular momentum approximation, we further obtain the free energy as
\begin{equation}
\begin{aligned}
F=\frac{4 \pi  C \left(10 \pi ^2 \mathcal{J}^2+\mathcal{S}^2\right)-\mathcal{S}^3}{4 \pi  \sqrt{C \mathcal{S}^3 \mathcal{V}}}+\mathcal{O}(\J^2).
\end{aligned}
\end{equation}
The temperature of the CFT is accurately given by \eq{cfttemp}; in the small angular momentum approximation, we get 
\begin{equation}\label{tempjvc}
\begin{aligned}
\T =\frac{4 \pi  C \left(\mathcal{S}^2-6 \pi ^2 \mathcal{J}^2\right)+3 \mathcal{S}^3}{4 \pi  \sqrt{C \mathcal{S}^5 \mathcal{V}}}+\mathcal{O}(\J^2).
\end{aligned}
\end{equation}

\begin{figure}[tbp!]
\begin{center}
\includegraphics[width=74mm,angle=0]{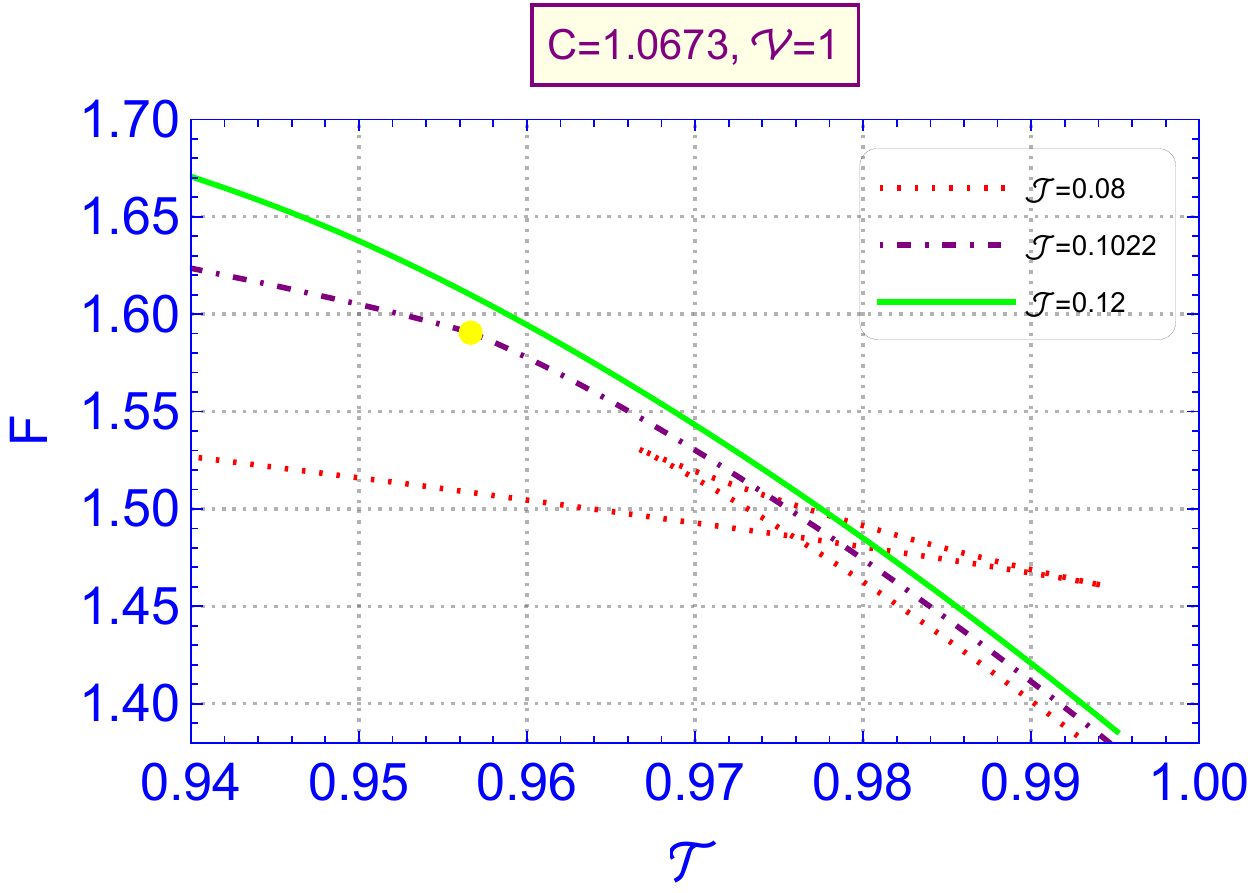}
\includegraphics[width=70mm,angle=0]{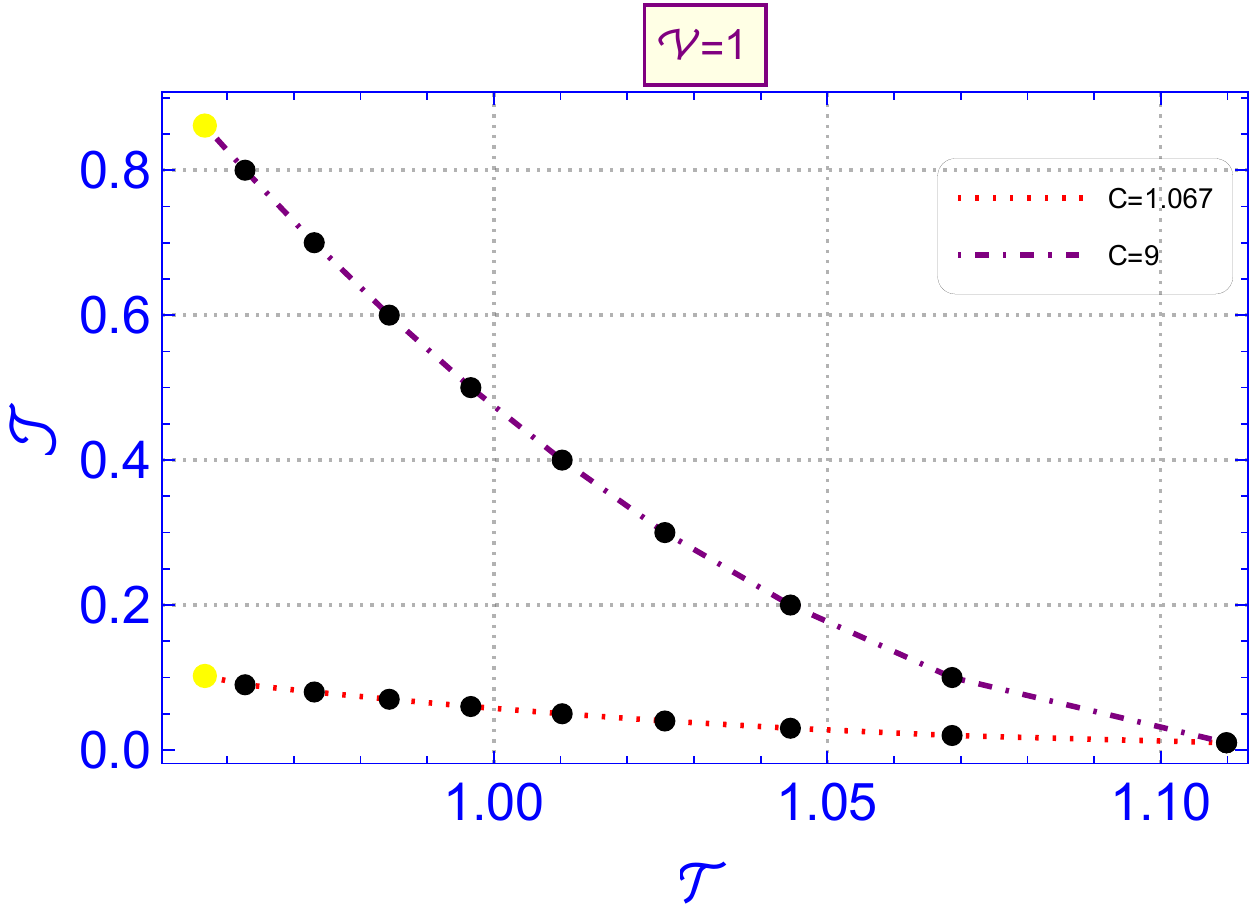}
\end{center}
\vspace{-5mm}
 \caption {Left: variations of the Helmholtz free energy $F$ with respect to the temperature $\T$ with the critical angular momentum $\J=0.1022$ in the fixed $(\J, \V, C)$ canonical ensemble. The yellow point is the critical point. Right: Coexistence lines indicating the variation of the coexistence temperature with respect to the angular momentum. The lines end at yellow critical points.}
 \label{fig1}
\end{figure}

Employing the free energy \eq{helfrc} and the temperature \eq{cfttemp}, we plot \fig{fig1}, where the behaviors of the free energy are shown for different values of angular momentum, and  the coexistence curves are given for different given values of the central charges. In the left diagram of \fig{fig1}, we just keep the $C, \V$ fixed. We can find that there is a critical point $\J=\J_c$ in the free energy curve. There, the CFT state undergoes a second-order phase transition, with no jump of the entropy $\S$ for the CFT from the low entropy per degree of freedom state to the higher entropy per degree of freedom state. (As $C$ is fixed, we can explain $S/C$ as the entropy per degree of freedom.) If $\J<\J_c$, we see that there is a ``swallowtail" shape displayed by the free energy curve, which means that there is a first-order phase transition at a self-intersection point for the CFT from a state with low entropy per degree of freedom to a state with high entropy per degree of freedom. If $\J>\J_c$, we cannot see different phases for the CFT states. It is evident that the behavior of the free energy in this fixed $(\J, \mathcal{V}, C)$ ensemble is similar to the one for the charged AdS black holes shown in \cite{Cong:2021fnf}.  

In the right diagram of \fig{fig1}, we show the coexistence line of the CFT states with low entropy  and high entropy for different values of $C$.  The low entropy CFT states lie below the coexistence line and the high entropy states lie over this line. The coexistence temperature $\T$ of those two kinds of states increases with the decreasing $\J$ for a given central charge $C$. We denote the critical points of the phase transition by yellow points on the curves. Beyond those points, there are no distinct phases for the CFT. One thing we want to note is that  the coexistence temperature of the CFT states cannot be zero. Indeed, by the approximated expression for the critical point given below in \eq{crip}, we see that this can happen only in a CFT with infinite volume $\V$.

\begin{figure}[tbp!]
\begin{center}
\includegraphics[width=75mm,angle=0]{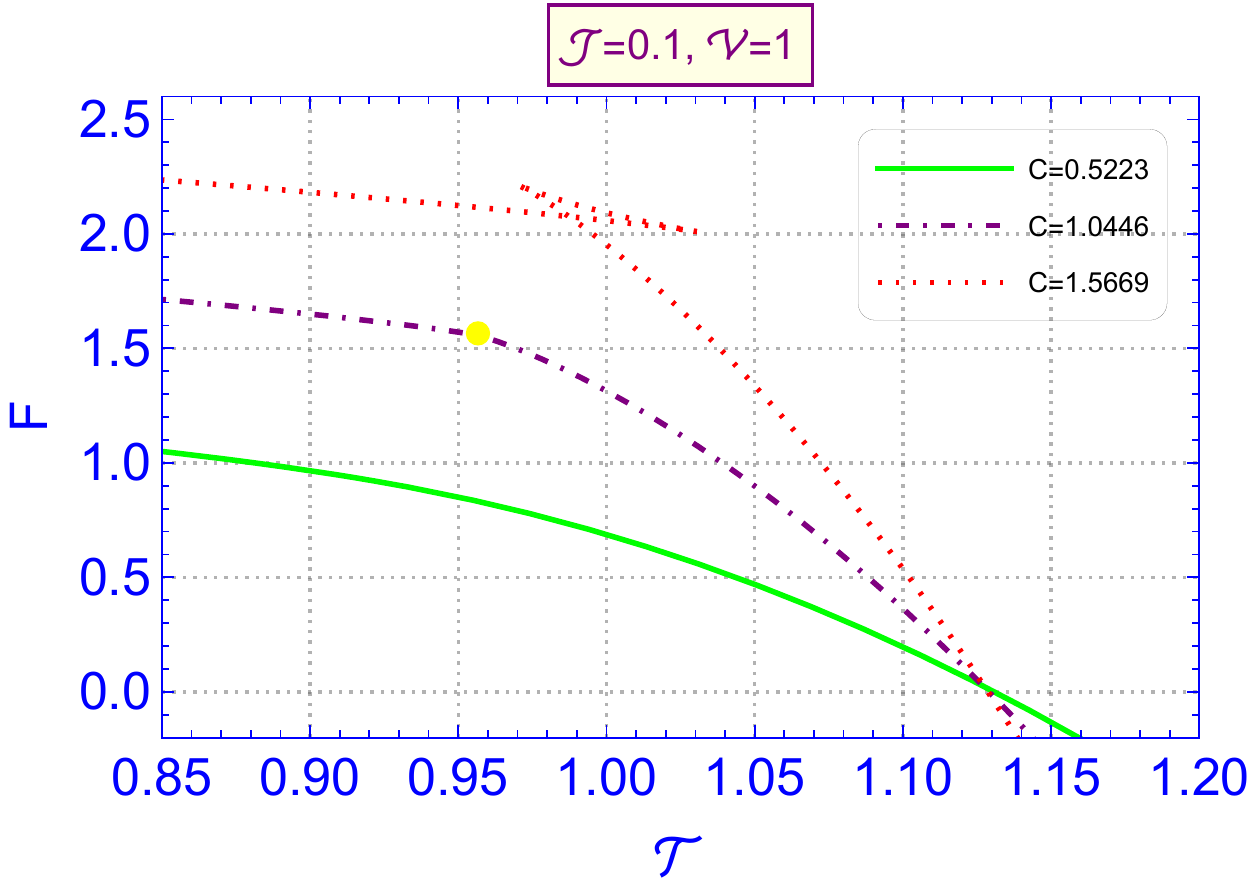}
\includegraphics[width=73mm,angle=0]{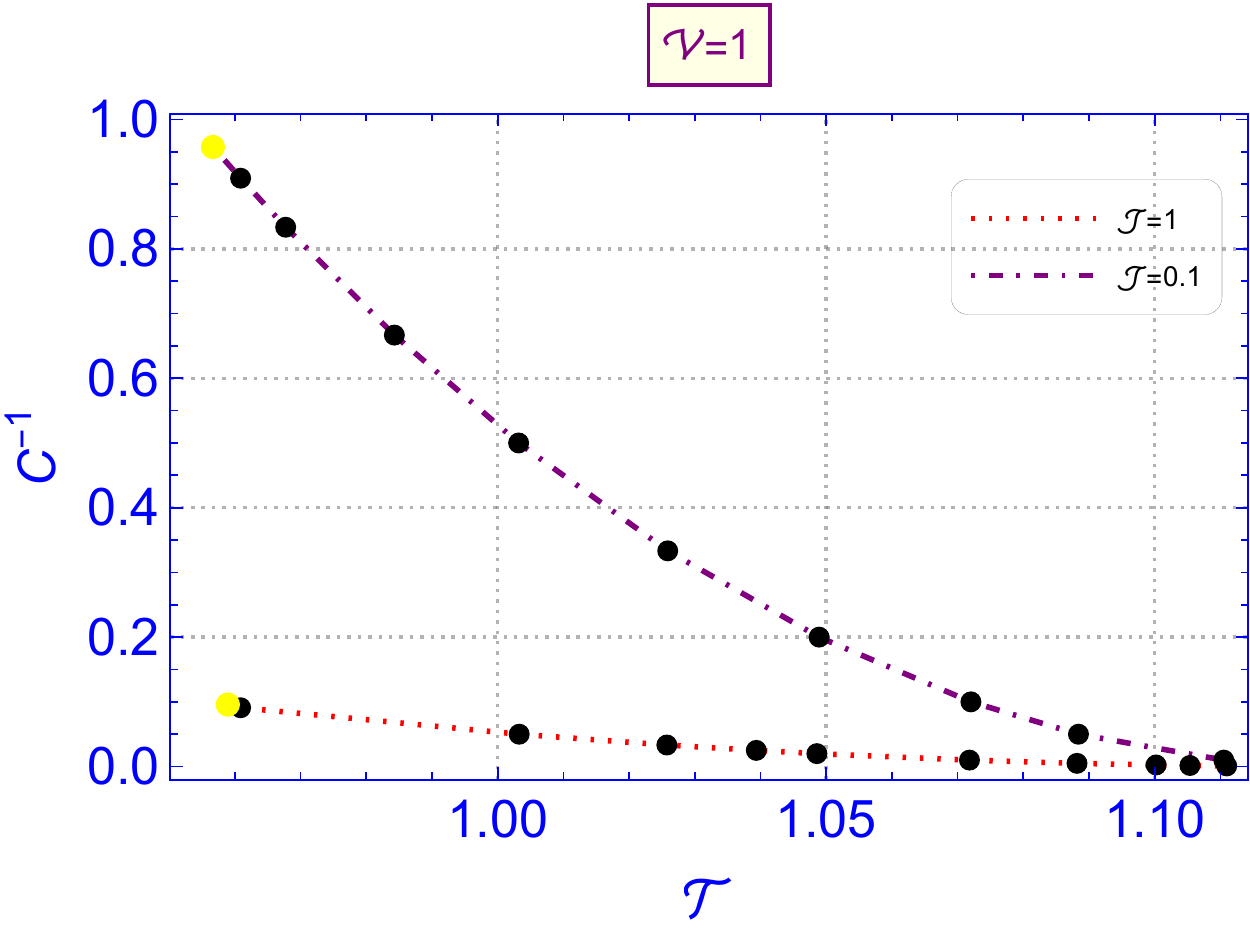}
\end{center}
\vspace{-5mm}
 \caption {Left: variations of the Helmholtz free energy $F$ with respect to the temperature $\T$ with the critical central charge $C=1.0446$ in the fixed $(\J, \V, C)$ canonical ensemble. The yellow point is the critical point. Right: Coexistence lines indicating the variation of the coexistence temperature with respect to the inverse central charge. The lines end at yellow critical points.}
 \label{fig3}
\end{figure}

In the left diagram of \fig{fig3}, we show the behavior of the free energy with different given central charges, $C<C_c, C=C_c, C>C_c$.  Qualitatively, we see that the free energy also shows swallowtail and kink behaviors just as the cases in \fig{fig1}. However, some more physics can be touched by further inspection of the critical values of the central charge. As we see, the first-order phase transition at the self-intersection point can happen only in the case $C>C_c$ while it is the $J<J_c$ for the case in \fig{fig1}. Moreover, if we plot the coexistence curve showing variation of the coexistence central charge with respect to the temperature, we will find that the coexistence temperature will increase with the increasing of the central charge. Besides, after comparing \fig{fig3} with \fig {fig0}, one will find that the central charge criticality of the CFT states is consistent with the central charge criticality of the dual rotating Kerr-AdS black hole.

To obtain explicitly the information on the critical points, we have to work in the small angular momentum configuration, where the temperature has been given by \eq{tempjvc}. Through the relations
\begin{equation}\label{crit}
\left(\frac{\partial \mathcal{T}}{\partial \mathcal{S}}\right)_{C,\mathcal{J},\V}=0=\left(\frac{\partial^2 \mathcal{T}}{\partial \mathcal{S}^2}\right)_{C,\mathcal{J},\V} ,
\end{equation}
we can find the approximated expressions of the critical points, i.e., the kink points in Figs. \ref{fig1} and \ref{fig3}, as
\begin{equation}\label{crip}
\begin{aligned}
\mathcal{S}_c&=3 \sqrt{10} \pi  \mathcal{J},\quad C_c=\frac{27}{4} \sqrt{\frac{5}{2}} \mathcal{J},\quad \mathcal{T}_c=\frac{6}{5} \sqrt{\frac{2}{\pi  \mathcal{V}}},\quad p_c=28 \sqrt{\frac{\pi }{5}} \mathcal{J} \mathcal{V}^{-3/2},\\ \mu_{c}&=-\frac{8}{15} \sqrt{\frac{2 \pi }{\mathcal{V}}},\quad \Omega_{c}=2 \sqrt{\frac{\pi }{5 \mathcal{V}}}.
\end{aligned}
\end{equation}
It is interesting that the critical temperature $\T_c$, pressure $p_c$, chemical potential $\mu_{c}$, and angular velocity $\Omega_{c}$ are dependent on the volume of the CFT while the critical entropy $\S_c$ and  central charge $C_c$ only depend on the angular momentum. As was stated above, we can have $\T_c\to 0$ only in the case of $\V\to \infty$.

Just like the charged AdS black hole case studied in \cite{Cong:2021fnf}, we here also note that the CFT states dual to the rotating AdS black holes cannot be analogous simply  with the Van der Waals fluids, typically as the coexistence curves in the $\J-\T$ and $C^{-1}-\T$ planes have negative slope while the coexistence curves in the pressure-temperature plane have positive slopes.

Another quite important phenomenon is that there is no $p-\V$ criticality phenomenon for the CFT states dual to (charged) rotating AdS black holes, as it is evident from \eq{mass3} and \eq{eoseq}, as we have $p\propto \sqrt{\V}$ for certain $\J, \mathcal{V}, C, \S$. This is quite different from the $P-V$ criticality of the Kerr AdS black hole \cite{Gunasekaran:2012dq}.

\subsection{Critical exponents}
We will study the scaling behavior of the macroscopic thermodynamic variables near the critical point of the CFT states. 
In the fixed $(\mathcal{J}, \mathcal{V}, C)$ ensemble, the critical point, including its approximated value \eq{crip} in the small angular momentum $\J$ regime, was  obtained in the $\T-\S$ plane above. Besides, in the first law \eq{affir} for the CFT, we can see another two planes, $C-\mu$ and $\mathcal{J}-\Omega$, that show criticality behaviors. However, as has been pointed out, there is no $p-\V$ criticality for the CFT states, this can be understood as the scaling property of the CFT volume $\V$ is not like the bulk volume $V$. The critical point \eq{crip} in the $C-\mu$ and $\mathcal{J}-\Omega$ planes can be obtained by relations similar to \eq{crit} for the $\T-\S$ conjugate pair.

We are to calculate the critical exponents of the phase transition for the CFT in the fixed $(\mathcal{J}, \mathcal{V}, C)$ ensemble and in the $C-\mu$ plane. A similar analysis can be conducted  in the $\J-\Omega$ plane and the same results can be obtained. For definiteness, we will work in the small angular momentum regime to proceed analytically.
 
Referring to \cite{Gunasekaran:2012dq}, we first define some dimensionless reduced variables as
\begin{equation}\label{cetcp}
t \equiv \frac{\mathcal{T}}{\mathcal{T}_c}-1, \quad \chi \equiv \frac{C}{C_c}, \quad \psi \equiv \frac{\mu}{\mu_c}-1.
\end{equation}
Then we can describe the behavior of some physical quantities, typically the specific heat $\mathcal{C}_{\mathcal{J}, \mathcal{V}, \mu}$, the order parameter $\eta \equiv \mu_h-\mu_l$ with $\mu_h$ $\left(\mu_l\right)$ being the chemical potential for the low(high) entropy states, the isothermal compressibility-like quantity $\kappa_\T$    at the vicinity of the critical point, respectively as
\begin{equation}\label{matc}
\mathcal{C}_{\mathcal{J}, \mathcal{V}, \mu} \equiv \mathcal{T}\left(\frac{\partial S}{\partial \mathcal{T}}\right)_{\mathcal{J}, \mathcal{V}, \mu} \sim|t|^{-\alpha},
\end{equation}
\begin{equation}
\eta \equiv \mu_h-\mu_l \sim|t|^\beta,
\end{equation}
\begin{equation}
\kappa_\mathcal{T} \equiv-\frac{1}{\mu}\left(\frac{\partial \mu}{\partial C}\right)_{\mathcal{T}, \J, \mathcal{V}} \sim|t|^{-\gamma}.
\end{equation}
Besides, near the critical point along the isotherm $\mathcal{T}=\mathcal{T}_{\text {crit }}$, the central charge behaves as
\begin{equation}
\left|C-C_c\right| \sim\left|\mu-\mu_c\right|^\delta,
\end{equation}
which singles out another critical exponent $\delta$.

To proceed analytically, we expand all quantities to $\O\left(J^2\right)$ in what follows. Though the exact critical point will shift from the one we get in \eq{crip}, it is unlikely that the critical exponents will change, as it was previously argued by \cite{Altamirano:2014tva} for the ones of the bulk Kerr-AdS black hole. With this prescription, we have the CFT energy corresponding to the Kerr-AdS black hole as
\begin{equation}
E=\frac{4 \pi  C \left(2 \pi ^2 \J^2+\S^2\right)+\S^3}{2 \pi  \sqrt{C \S^3 \mathcal{V}}}+\O(\J^2).
\end{equation}
Then the temperature $\T$, chemical potential $\mu$ of the CFT are \footnote{Note that the temperature obtained in this way (i.e., derived from the approximated CFT energy) is the same with \eq{tempjvc} (approximated from accurate temperature). So does it for other quantities.}
\begin{equation}\label{cett}
\T=\frac{4 \pi  C \left(\S^2-6 \pi ^2 \J^2\right)+3 \S^3}{4 \pi  \sqrt{C \S^5 \mathcal{V}}}+\O(\J^2),
\end{equation}
\begin{equation}\label{cemm}
\mu=\frac{4 \pi  C \left(2 \pi ^2 \J^2+\S^2\right)-\S^3}{4 \pi  C \sqrt{C \S^3 \mathcal{V}}}+\O(\J^2).
\end{equation}
In what follows, for simplicity, we neglect the higher order term $\O(\J^2)$. By definition, we then get
\begin{equation}
\mathcal{C}_{\mathcal{J}, \mathcal{V}, \mu}=-\frac{\mathcal{S} \left(4 \pi  C \left(\mathcal{S}^2-6 \pi ^2 \mathcal{J}^2\right)+3 \mathcal{S}^3\right) \left(3 \mathcal{S}^3-4 \pi  C \left(2 \pi ^2 \mathcal{J}^2+\mathcal{S}^2\right)\right)}{32 \pi ^3 C \mathcal{J}^2 \left(4 \pi  C \left(6 \pi ^2 \mathcal{J}^2+\mathcal{S}^2\right)-3 \mathcal{S}^3\right)},
\end{equation}
which, at the critical point \eq{crip}, reduces to be
\begin{equation}
\left(\mathcal{C}_{\mathcal{J}, \mathcal{V}, \mu}\right)_c=48 \sqrt{10} \pi  \J.
\end{equation}
We can see that the specific heat of the CFT state is not divergent at the critical point, yielding $\alpha=0$ according to \eq{matc}.

Using \eq{crip} and \eq{cett}, we have
\begin{equation}\label{cechi}
\chi=\frac{\sqrt{2} \left(18 \pi ^2 \mathcal{J}^2 \mathcal{S}^3+2 \sqrt{\pi }\S^4\T \sqrt{ \mathcal{V} \left(18 \pi ^2 \mathcal{J}^2+\mathcal{S}^2 \left(\pi  \mathcal{T}^2 \mathcal{V}-3\right)\right)}+\mathcal{S}^5 \left(2 \pi  \mathcal{T}^2 \mathcal{V}-3\right)\right)}{27\sqrt{5} \pi  \mathcal{J} \left(\mathcal{S}^2-6 \pi ^2 \mathcal{J}^2\right)^2}.
\end{equation}
Furthermore, with the help of \eq{crip} and \eq{cemm}, we get
\begin{equation}\label{cepsi}
\psi=\frac{15 \left(4 \pi  C \left(2 \pi ^2 \mathcal{J}^2+\mathcal{S}^2\right)-\mathcal{S}^3\right)}{32 \sqrt{2} \pi ^{3/2} (C \mathcal{S})^{3/2}}-1.
\end{equation}
Then combing \eq{crip}, \eq{cetcp},  \eq{cechi}, and \eq{cepsi}, we obtain
\begin{equation}\label{chitpsi}
\begin{aligned}
\chi&=1+12t-540 t \psi +7290 \psi ^3+\O\left(t \psi^2, \psi^4\right)\\
&\equiv1+\bar{A}t+\bar{B}t\psi+\bar{C}\psi^3+\O\left(t \psi^2, \psi^4\right),
\end{aligned}
\end{equation}
where, for later convenience, we have used some algebraic quantities $\bar{A}, \bar{B}, \bar{C}$. 

For the first-order phase transition near the critical point in the $C-\mu$ plane, we have the Maxwell's equal area law
\begin{equation}\label{meal}
\int_{\psi_l}^{\psi_h}\psi dC=0,
\end{equation}
where 
\begin{align}
\psi_h\equiv\frac{\mu_h}{ \mu_c}-1,\quad
\psi_l\equiv\frac{\mu_l}{ \mu_c}-1,
\end{align}
and on the isotherm, we have 
\begin{equation}
d C=C_c\left(\bar{B} t+3 \bar{C} \psi ^2\right) d \psi.
\end{equation}
Moreover, in the first-order phase transition near the critical point, we also have invariable $\chi$ (just like the invariable $P$ for the bulk Kerr-AdS black hole). This gives
\begin{equation}\label{chibb}
\begin{aligned}
\chi=1+\bar{A}t+\bar{B}t\psi_h+\bar{C}\psi^3_h=1+\bar{A}t+\bar{B}t\psi_l+\bar{C}\psi^3_l.
\end{aligned}
\end{equation}
\eq{meal} and \eq{chibb} together yield a  unique non-trivial solution as
\begin{equation}
\psi_h=-\psi_l=\sqrt{\frac{-\bar{B} t}{\bar{C}}},
\end{equation}
which immediately gives
\begin{equation}
\eta=\mu_{c}\left(\psi_h-\psi_l\right)=2\mu_{c}\sqrt{\frac{-\bar{B} t}{\bar{C}}}.
\end{equation}
Thus we get to know the exponent $\beta=1 / 2$.

Employing \eq{chitpsi}, it is straightforwardly to obtain
\begin{equation}
\left(\frac{\partial \mu}{\partial C}\right)_{\T, \J, \mathcal{V}}=\frac{\mu_c}{\bar{B}C_c t}+\O(\psi^2),
\end{equation}
which further implies
\begin{equation}
\kappa_T \sim \frac{1}{\bar{B}C_c t}.
\end{equation}
So we have $\gamma=1 $. Again as long as setting $t=0$ in \eq{chitpsi}, we can obtain
\begin{equation}
\chi=1+\bar{C} \psi^3,
\end{equation}
which immediately shows $\delta=3$. 

As we can see, the critical exponents, exponents of scaling laws that describe the behavior of the CFT states in the vicinity of the critical point, are the same as that of the rotating black hole previously found in  \cite{Gunasekaran:2012dq}. This justifies that the critical exponents of the CFT criticality  belong to the universality class predicted by the mean field theory.

\section{Phase transition of CFT in grand canonical ensemble}\label{oten}

We will now study the thermodynamic phase structure of the dual CFT in the grand canonical ensemble where $(\Omega, \mathcal{V}, C)$ are fixed.  From our above analysis for the CFT in the canonical ensemble with fixed $(\J, \mathcal{V}, C)$, we found that the role of $\J$ in the CFT is much alike the role of $J$ in the bulk, so we wonder whether it is also so for $\Omega_b$ and $\Omega$. Note that the Hawking-Page phase transition of the KN-AdS black hole was studied in \cite{Caldarelli:1999xj}.  The thermodynamic potential associated with this ensemble (so that all other thermodynamic state functions can be derived from it) is the Gibbs free energy
\begin{equation}
W \equiv E-\T \S-\Omega \J=\mu C.
\end{equation}
Here we have used the notation convention of \cite{Cong:2021fnf} for the Gibbs free energy that is  used for the charged AdS black hole to make it convenient for the reader to compare the results.

\begin{figure}[tbp!]
\begin{center}
\includegraphics[width=75mm,angle=0]{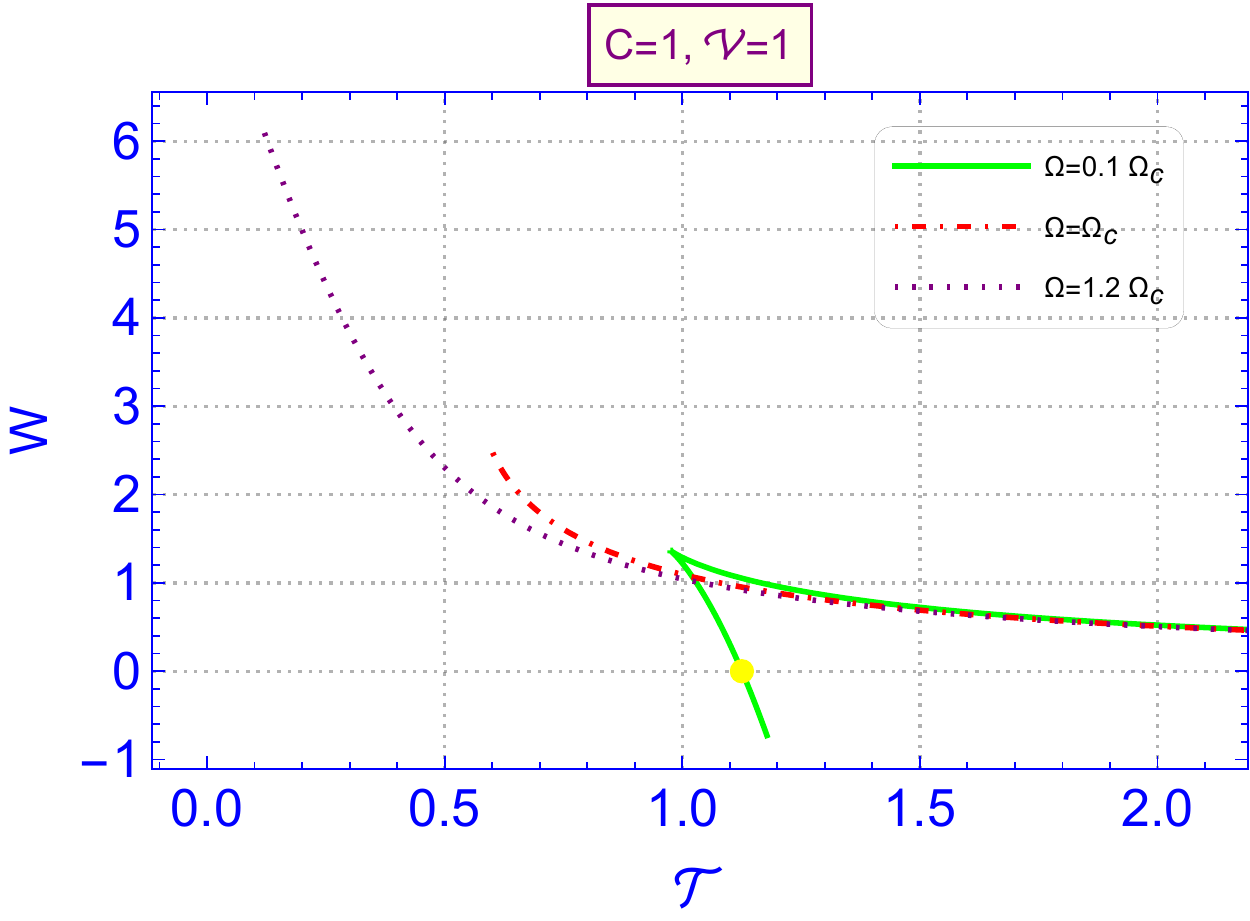}
\includegraphics[width=75mm,angle=0]{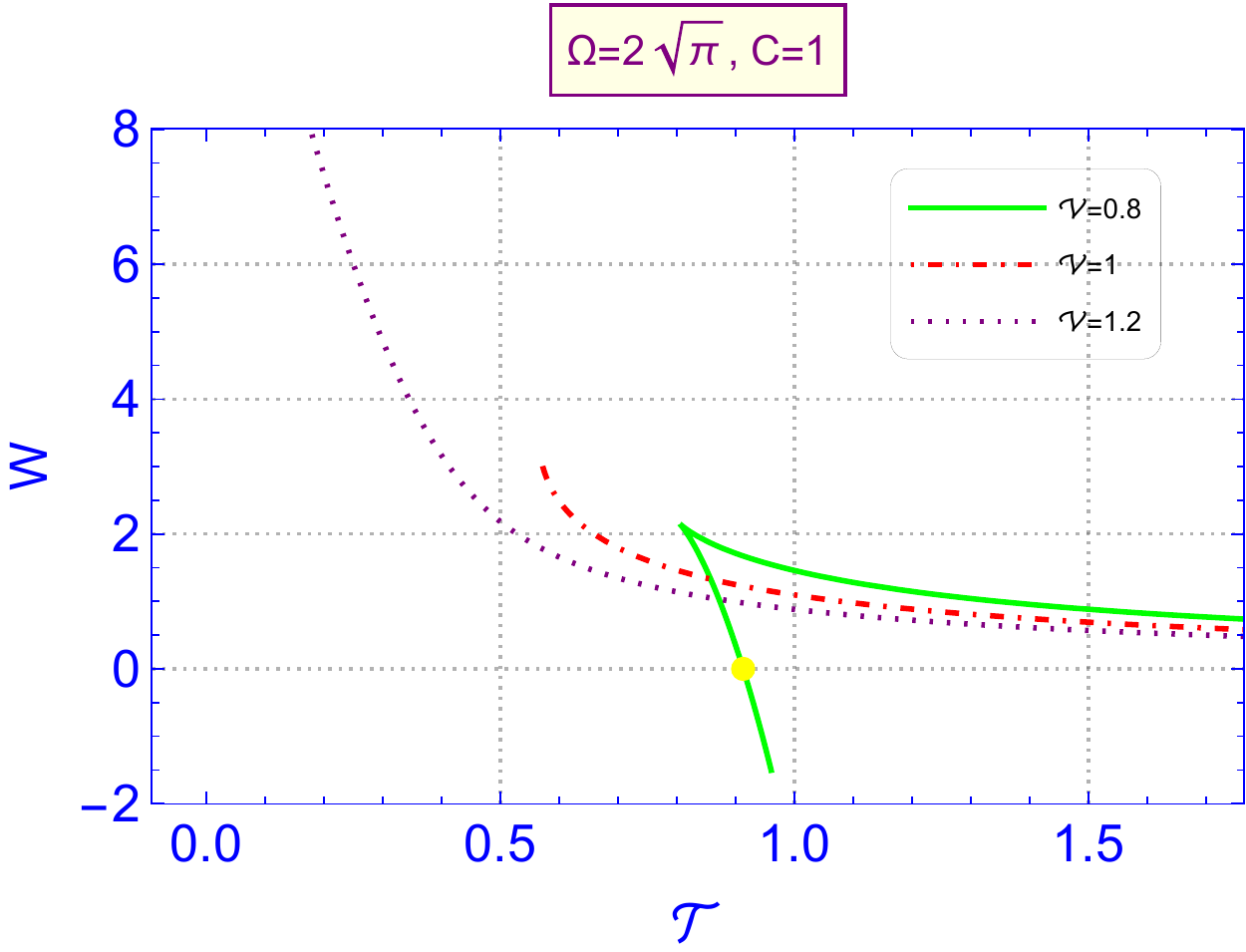}
\end{center}
\vspace{-5mm}
 \caption {The variations of the Gibbs free energy $W$ with respect to the temperature $\T$ in the fixed $(\Omega, \mathcal{V}, C)$ grand canonical ensemble. The (de)confinement phase transitions take place at the yellow point where $W$ changes its sign.}
 \label{fig4}
\end{figure}

First, from the angular momentum
\begin{equation}
\begin{aligned}
\Omega=\frac{8 \pi ^2 C \mathcal{J} (4 \pi  C+\mathcal{S})}{\sqrt{C \mathcal{S} \mathcal{V} (4 \pi  C+\mathcal{S}) \left(4 \pi  C \left(4 \pi ^2 \mathcal{J}^2+\mathcal{S}^2\right)+\mathcal{S}^3\right)}},
\end{aligned}
\end{equation}
we obtain
\begin{equation}
\J=\frac{\mathcal{S}^{3/2} \Omega  \sqrt{(4 \pi  C+\mathcal{S})\V}}{\sqrt{256 \pi ^5 C^2-16 \pi ^3 C \mathcal{S} \mathcal{V} \Omega ^2+64 \pi ^4 C \mathcal{S}}}.
\end{equation}
Then we can easily get the Gibbs free energy and the temperature in terms of $\Omega$ individually as
\begin{equation}
\begin{aligned}
W&=\frac{16 \pi ^2 C^2 \left(4 \pi ^2 \mathcal{J}^2+\mathcal{S}^2\right)-\mathcal{S}^4}{4 \pi  \sqrt{C \mathcal{S} \mathcal{V} (4 \pi  C+\mathcal{S}) \left(4 \pi  C \left(4 \pi ^2 \mathcal{J}^2+\mathcal{S}^2\right)+\mathcal{S}^3\right)}}\\&=\frac{64 \pi ^3 C^2+\mathcal{S}^2 \left(\mathcal{V} \Omega ^2-4 \pi \right)}{8 \pi ^{3/2} C \mathcal{S} \mathcal{V} (4 \pi  C+\mathcal{S})^2}\sqrt{\frac{C \mathcal{S}^3 \mathcal{V} (4 \pi  C+\mathcal{S})^3}{16 \pi ^2 C-\mathcal{S} \mathcal{V} \Omega ^2+4 \pi  \mathcal{S}}},
\end{aligned}
\end{equation}
\begin{equation}
\begin{aligned}
\T =\frac{64 \pi ^3 C^2+8 \pi  C \mathcal{S} \left(8 \pi -\mathcal{V} \Omega ^2\right)+3 \mathcal{S}^2 \left(4 \pi -\mathcal{V} \Omega ^2\right)}{8 \pi ^{3/2} C \mathcal{S}^3 \mathcal{V} (4 \pi  C+\mathcal{S})^2}\sqrt{\frac{C \mathcal{S}^5 \mathcal{V} (4 \pi  C+\mathcal{S})^3}{16 \pi ^2 C-\mathcal{S} \mathcal{V} \Omega ^2+4 \pi  \mathcal{S}}}.
\end{aligned}
\end{equation}

With these quantities at hand, we show the variation of $W$ in terms of $T$ in \fig{fig4}. For certain values of $\Omega$, we can see that the Gibbs free energy exhibits either single-valued behavior or a bifurcation point from which two branches emerge. The cusp can be calculated from the temperature as 
\begin{equation}
\left(\frac{\partial \T}{\partial \S}\right)_{\Omega,\mathcal{V},C}=0,
\end{equation}
which gives the corresponding entropy and the temperature of the cusp point as $\S_{\mathrm{cusp}},\,\T(\S_{\mathrm{cusp}})$, whose explicit forms are not given here as they are lengthy and lack evident inspiration. However, we do know from the equation that if 
\begin{equation}
4 \pi -\mathcal{V} \Omega ^2\leqslant 0, \,\mathrm{i.e.,}\, \Omega\geqslant\Omega_c\equiv \frac{2 \sqrt{ \pi }}{\sqrt{\mathcal{V}}},
\end{equation}
we cannot have the cusp at the physically allowed region $\S>0$. In other words, if the above condition holds true, a single-valued Gibbs free energy can be obtained.

\begin{figure}[tbp!]
\begin{center}
\includegraphics[width=85mm,angle=0]{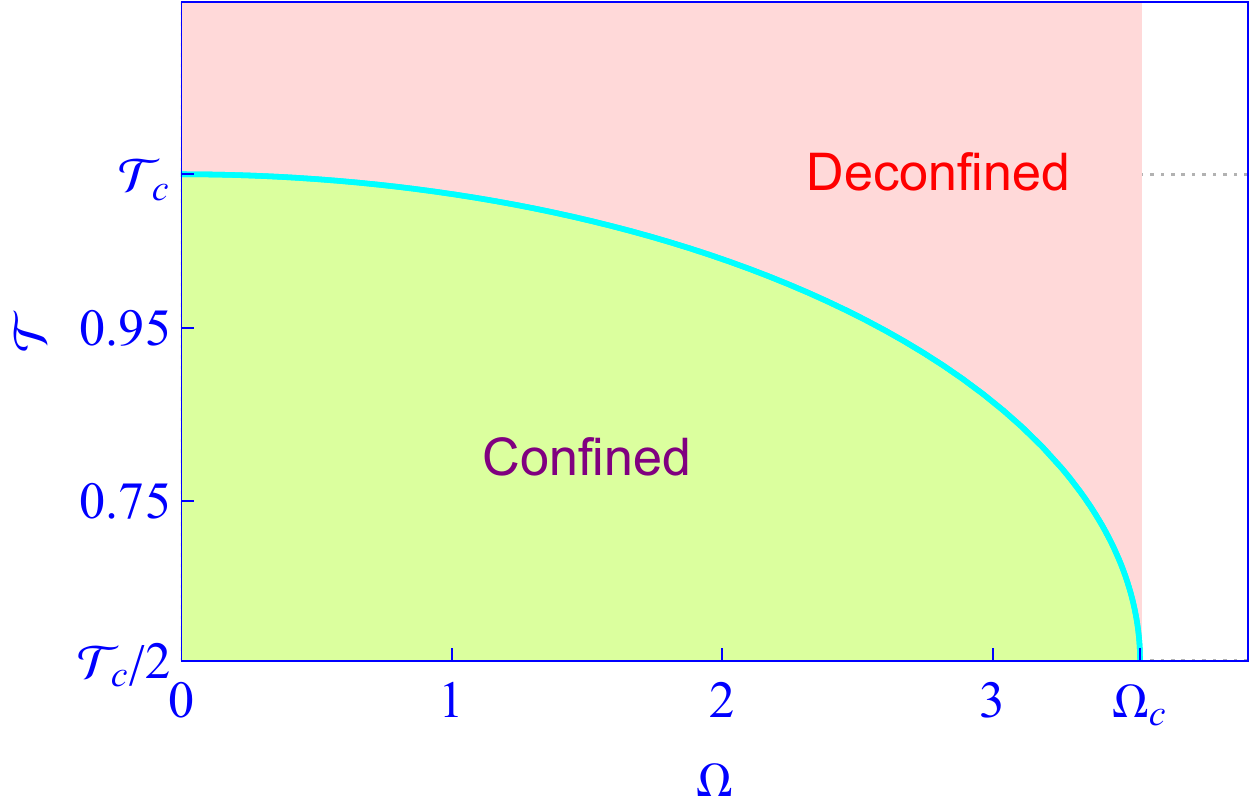}
\end{center}
\vspace{-5mm}
 \caption {The variations of the temperature $\T$ with respect to the angular velocity $\Omega$ for the CFT states in the fixed $(\Omega, \mathcal{V}, C)$ grand canonical ensemble. The (de)confinement phase transitions take place on the cyan coexistence line, below which is the confined state and above which is the deconfined state.}
 \label{fig6}
\end{figure}

From \fig{fig4}, we can also see that there is a point where the Gibbs free energy changes its sign. This does mean a phase transition. Specifically, we know that a system with lower free energy  is a thermodynamically preferred configuration. So here the physical explanation is: when the temperature of the CFT states increases from zero, the confined state dominates this grand canonical ensemble until the turning point where $W$ changes its sign from positive to negative; there the system has a first-order phase transition and after that, the system is instead dominated by the large entropy deconfined state. This is reminiscent of the  (generalized) Hawking-Page phase transition that was discovered for the Schwarzschild-AdS black holes, RN-AdS black holes, and KN-AdS black holes. Quantitively, we can determine the phase transition point by $W=0$, which yields a coexistence line of the (de)confinement first-order phase transition as
\begin{equation}
\left(\frac{2\T}{\T_c}-1\right)^2+\left(\frac{\Omega }{\Omega_c}\right)^2=1,\, \frac{\T_c}{2}< \T\leqslant \T_c,\, 0\leqslant\Omega< \Omega_c,
\end{equation}
where we have defined the temperature $\T_c=2/\sqrt{\pi\V}$ for the (de)confinement phase transition dual to the Hawking-Page phase transition of a Schwarzschild-AdS black hole. \fig{fig6} displays the coexistence line of the (de)confinement phase transition, showing both the confinement and deconfinement regimes. As we approach the limit of infinite volume ($\V\to\infty$), we can see that the (de)confinement temperature $\T\to 0^+$ for  $\Omega\to\Omega_c^-$. Note $T_c$ here is equivalent to the one obtained in \cite{Cong:2021jgb}. It's worth noting that if $\Omega\geqslant\Omega_c$, there can be no (de)confinement phase transition. This result aligns with a recent discovery by \cite{Ahmed:2023dnh}, in which the superradiant instability of the corresponding bulk black hole were also considered when $\Omega=\Omega_c$.

\section{Concluding remarks}

In this paper, we first presented explicit  mass/energy formulas \eq{mass1t}, \eq{mass2}, \eq{mass3} for the extended thermodynamics, mixed thermodynamics, and CFT thermodynamics for the KN-AdS black hole and its holographic dual CFT. Thermodynamic state functions can be directly derived from the thermodynamic first law by employing the mass/energy formulas. Note that the mass formula \eq{mass1} can be obtained by restoring the Newton's constant from the one given in \cite{Caldarelli:1999xj}, or in other words, it can be obtained by first solving $a, l, r_+$ from $J, P, S$ using \eq{mjq}, \eq{preb}, and \eq{entb} before substituting them into \eq{main}. This happens to be a privilege for the four-dimensional KN-AdS black hole but is not effective for higher-dimensional Kerr-AdS black holes. That is to say, the extension of these mass/energy formulas to the higher-dimensional case seems impossible.

Then we investigated the phase structures of holographic dual CFT thermal states for Kerr-AdS black holes in both the canonical and grand canonical ensembles. In the canonical ensemble with fixed $(\J, \V, C)$, we found angular momentum criticality and central charge criticality for the CFT states. The CFT undergoes a first-order phase transition from a small $\S/C$ or $\S/\J$ state to a large $\S/C$ or $\S/\J$ state. However, coexistence analysis showed that these CFT states cannot be analogous to the Van der Waals fluid. Nonetheless, the critical exponents of the CFT states at the critical point of the second-order phase transition are uniformly similar to those of the Van der Waals fluid. This scheme-independent characteristics of the critical exponents suggest that it can be derived without detailed knowledge of the microscopic properties of the systems and the universality can be understood with the renormalization group. Note that the critical exponents are analytically derived in small angular momentum approximation. We believe that these critical exponents will remain unchanged even without the small angular momentum approximation. In the grand canonical ensemble with fixed $(\Omega, \V, C)$, we found a first-order (de)confinement phase transition. Coexistence analysis shows that the phase transition of the CFT states can occur at zero temperature with the angular velocity of the CFT dependent solely on the CFT volume. Our findings for both canonical and grand canonical ensembles are consistent with those of charged AdS black holes as presented in \cite{Cong:2021jgb}.

In \cite{Cong:2021jgb}, a comprehensive analysis was conducted on the phase behavior of all possible ensembles for the charged AdS black hole dual CFT. We did not replicate this analysis as our investigations of the fixed $(\J, \V, C)$ and $(\Omega, \V, C)$ ensembles indicated that $\{\J,\Omega\}$ play a similar role to $\{\Q,\Phi\}$. Through calculations, we found the existence of a zeroth-order phase transition of the CFT in the fixed $(\J, \V, \mu)$ ensemble, and also discovered that in the small $\J$ approximation to $\O(\J^2)$, there is no zeroth-order phase transition as $(d\T/d\S)_{\J, \V, \mu}>0$ within the physically allowed parameter range, ensuring that the CFT states' specific heat remains always positive. Although there are additional ensembles when considering CFT states dual to KN-AdS black holes, the interplay between $\{\J,\Omega\}$ and $\{\Q,\Phi\}$ is likely to be complex. Fortunately, the energy formulas we have provided above will likely simplify related calculations.

\acknowledgments
 MZ is supported by the National Natural Science Foundation of China with Grant No. 12005080.  JJ  is supported by the National Natural Science Foundation of China with Grant No. 210510101, the Guangdong Basic and Applied Research Foundation with Grant No. 217200003, and the Talents Introduction Foundation of Beijing Normal University with Grant No. 310432102.

\bibliographystyle{jhep}
\bibliography{refs}
\end{document}